\documentclass{article}[12pt]
\usepackage{color}
\usepackage{booktabs}
\usepackage[pdftex]{graphicx}   
\usepackage{amsmath, amssymb}
\usepackage{stmaryrd}
\begin{document}
\title{Construction of Superposition States of Energy Eigenstates via Classically Emulated Digital Quantum Simulation: The Hydrogen Molecule as an Example
 }
\author{Kazuto Oshima\thanks{E-mail: kooshima@gunma-ct.ac.jp}    \\ \\
\sl National Institute of Technology, Gunma College, Maebashi 371-8530, Japan}
\date{}
\maketitle
\begin{abstract}
We construct superposition states of energy eigenstates of the hydrogen molecule using classically emulated digital quantum simulation.
We generate the ground state and excited states of the system via the twirling operation method,
and construct superposition states of the ground state and an excited state of the system by applying a controlled excitation unitary transformation
on the ground state with one ancillary qubit as the control.
To verify the correctness of the resulting superposition state, we calculate matrix elements of several physical observables.
\end{abstract} 

\section{Introduction}
Understanding and simulating molecular energy eigenstates is a central challenge in quantum chemistry and quantum simulation. Conventional quantum algorithms such as the variational quantum eigensolver (VQE) \cite{Peruzzo}, the variational quantum deflation (VQD) \cite{Higgott}, and the quantum phase estimation (QPE) \cite{Kitaev} have been widely used to address this problem.
A novel approach has recently been proposed for generating energy eigenstates of the hydrogen molecule via classically emulated quantum simulation \cite{Oshima1}.
In the VQE framework, the generation of excited states typically requires additional techniques, such as the VQD.
In contrast, our approach enables a unified treatment of both ground and excited states within a single framework.
QPE inherently suffers from the difficulty of extracting quantum states with small amplitudes due to their negligible measurement probabilities.
In contrast, our approach allows for the extraction of states with initially small amplitude through a progressive amplification process.

The purpose of this paper is to demonstrate how superpositions of a quantum system's eigenstates can be constructed via classically emulated digital quantum simulation.
Since the hydrogen molecule serves as a benchmark system in quantum chemistry, it is used as an illustrative example.
We construct superposition states of energy eigenstates of the hydrogen molecule based on the previously introduced approach for generating energy eigenstates.
Superposition states comprising ground and excited states of the hydrogen molecule are constructed by applying quantum circuits incorporating a unitary operator designed to map the ground state
to an excited state.
To understand the dynamics of a quantum system, it is essential to properly construct and analyze superpositions of energy eigenstates.
Beyond its relevance to quantum chemistry, constructing superpositions of energy eigenstates also holds significance from the perspective of quantum information science.

In Section 2, we describe the twirling operations,  which play a crucial role in generating energy eigenstates of a Hamiltonian.
In Section 3, we describe two types of superposition states constructed from energy eigenstates.
Sections 4 and 5 present simulation results for a single-qubit model and the hydrogen molecule, respectively. Section 6 concludes with a summary and discussion.

We performed simulations using a 13th Gen Intel Core i7 1.70GHz personal computer.  Python 3.8.8 and IBM Qiskit 0.45.0 were used for the simulations.
The main simulation results can be reproduced using the code available in the repository at https://github.com/kazuto-oshima/.

\section{Twirling operations for generating energy eigenstates}
In this section we introduce twirling operations for generating energy eigenstates \cite{Oshima2}.    Given a Hamiltonian $H$ of a quantum system, the following sequence of operations applied to 
a physical quantum state $|\psi\rangle$ and an ancilla qubit initialized in state $|0\rangle=\left(\begin{array}{c} 1 \\ 0 \end{array}\right)$ is referred to as a single twirling operation, as illustrated in Fig. 1, 
\begin{eqnarray}
|\psi\rangle|0\rangle  \xrightarrow{H_{d}} |\psi\rangle{|0\rangle +|1\rangle \over \sqrt{2}}  \xrightarrow{c-U(\tau)}{1 \over \sqrt{2}}( |\psi\rangle |0\rangle+e^{-i\tau{H}} |\psi\rangle  |1\rangle) \nonumber \\
\xrightarrow {H_{d}} {1 \over 2}(1+e^{-i\tau{H}})|\psi\rangle|0\rangle+{1 \over 2}(1-e^{-i\tau{H}})|\psi\rangle|1\rangle,
\end{eqnarray}
where $H_{d}={1 \over \sqrt{2}}\left(\begin{array}{cc} 1 & 1 \\ 1 & -1 \end{array}\right)$ denotes the Hadamard transformation and  $|1\rangle=\left(\begin{array}{c} 0 \\ 1 \end{array}\right)$.
Let $\{|E_{j}\rangle\}_{\{j=0,1,2,\dots,n\}}$ be the energy eigenstates of the Hamiltonian $H$.   Consider the case in which the physical state $|\psi\rangle$ is expressed as $|\psi\rangle=\sum_{j=0}^{n}c_{j}|E_{j}\rangle$.   By postselecting the ancilla qubit in state $|0\rangle$ in the final expression of Eq. (1), the resulting unnormalized physical state is given by $\sum_{j=0}^{n}{1+e^{-i\tau{E_{j}}} \over 2}c_{j}|E_{j}\rangle$.  Therefore, the amplitude of the state $|E_{j}\rangle$ corresponding to the largest value of $|{1+e^{-i\tau{E_{j}}} \over 2}|$ is relatively enhanced.    By introducing a new qubit initialized in the $|0\rangle$ state and
repeating this operation, the physical state corresponding to the ancilla qubits being in $|0\rangle|0\rangle \cdots |0\rangle$ can approximate $|E_{j}\rangle$ more closely.   This allows us to extract a single energy eigenstate contained in the initial physical state $|\psi\rangle$. 

There are two distinct ways to repeat the twirling operation \cite{Oshima1}.   One approach is to repeatedly apply the twirling operation with fixed duration $\tau$.   The other approach is to  vary the duration $\tau$ adaptively in each twirling operation.  The former is called a fixed-duration twirling operation, which is used to
extract a specific energy eigenstate.    The latter is referred to as an adaptive twirling operation, which is primarily useful for improving the accuracy of approximating an energy eigenstate.  \\
\\

*** Fig. 1. ***\\

\section{Construction of superposition states of energy eigenstates}
Let us assume that an arbitrary energy eigenstate $|E_{i}\rangle$ (where $(i=0,1,2,\cdots)$) is generated via the twirling operations.  In this context,  $|E_{0}\rangle$ refers to the ground state, while $|E_{i}\rangle$ indicates the $i$-th excited state.
By performing measurements in suitably chosen bases,
one can obtain the representation of these states as components in the tensor product state of the standard basis.
We propose a method to construct a superposition state of the ground state $|E_{0}\rangle$ and the excited state $|E_{i}\rangle(i \ge 1)$.   In the following,  
we describe how to construct a superposition of the ground state and the first excited state.  However, our procedure is not limited to this pair;
it is applicable to any pair of energy eigenstates.   We define the following unitary operator $U_{10}$ that maps the ground state $|E_{0}\rangle$
to the first excited state $|E_{1}\rangle$\cite{Sugisaki}
\begin{equation}
U_{10}=|E_{1}\rangle\langle E_{0}|+|E_{0}\rangle\langle E_{1}|+\sum_{i=2}|E_{i}\rangle\langle E_{i}|.
\end{equation}
Given a ground state, a superposition of the ground state and the first excited state can be created by the following algorithm, represented by
the quantum circuit in Fig. 2.   The algorithm starts by applying a Hadamard transformation $H_{d}={1 \over \sqrt{2}}\left(\begin{array}{cc} 1 & 1 \\ 1 & -1 \end{array}\right)$ to the ancilla qubit initially in the state $|0\rangle=\left(\begin{array}{c} 1 \\ 0 \end{array}\right) $ in Fig. 2(Eq. (3)).   
The next step is to apply the controlled-unitary transformation c-$U_{10}$.  The unitary operator $U_{10}$ is conditionally applied to $|E_{0}\rangle$ when the ancilla qubit is in the state $|1\rangle=\left(\begin{array}{c} 0 \\ 1 \end{array}\right)$(Eq. (4)).   A subsequent Hadamard transformation on the ancilla qubit yields the quantum state in Eq. (5).
\begin{eqnarray}
|0\rangle|E_{0}\rangle &\overset{\text{$H_{d}$ $\otimes$ $I$}}{\longrightarrow}& {1 \over \sqrt{2}}(|0\rangle+|1\rangle)|E_{0}\rangle\\
& \overset{\text{c-$U_{10}$}} {\longrightarrow}&  {1 \over \sqrt{2}}(|0\rangle|E_{0}\rangle+|1\rangle|E_{1}\rangle)\\
& \overset{\text{$H_{d}$ $\otimes$ $I$}}{\longrightarrow}& {1 \over \sqrt{2}}|0\rangle{1 \over \sqrt{2}}(|E_{0}\rangle+|E_{1}\rangle)+ {1 \over \sqrt{2}}|1\rangle{1 \over \sqrt{2}}(|E_{0}\rangle-|E_{1}\rangle).
\end{eqnarray}
\\
*****Fig. 2.*****\\
\\
If we measure a physical observable $Q$ in this state, the results are as follows.  When the ancilla qubit is measured as $0$, the expectation value of $Q$
is given by
\begin{equation}
\langle Q \rangle_{0}={1 \over 2}(\langle E_{0}|Q|E_{0}\rangle+\langle E_{1}|Q|E_{1}\rangle+\langle E_{1}|Q|E_{0}\rangle+\langle E_{0}|Q|E_{1}\rangle).
\end{equation}
When the ancilla qubit is measured as $1$, the expectation value of $Q$
is given by
\begin{equation}
\langle Q \rangle_{1}={1 \over 2}(\langle E_{0}|Q|E_{0}\rangle+\langle E_{1}|Q|E_{1}\rangle-\langle E_{1}|Q|E_{0}\rangle-\langle E_{0}|Q|E_{1}\rangle).
\end{equation}
By taking the difference between Eqs. (6) and (7), we obtain
\begin{equation}
\langle Q \rangle_{0}-\langle Q \rangle_{1}=\langle E_{1}|Q|E_{0}\rangle+\langle E_{0}|Q|E_{1}\rangle.
\end{equation}
From Eq. (8), the real part of $\langle E_{1}|Q|E_{0}\rangle$ can be determined.

This time we consider a slightly different quantum circuit (Fig. 3).   The operator $S(i)$ performs the following phase transformation.
\begin{equation}
S(i)(\alpha|0\rangle+\beta|1\rangle)=\alpha|0\rangle+i\beta|1\rangle.
\end{equation}
\\
*****Fig. 3.*****\\
\\
The combined physical and ancilla state is now given by
\begin{equation}
|\Psi\rangle= {1 \over \sqrt{2}}|0\rangle{1 \over \sqrt{2}}(|E_{0}\rangle+i|E_{1}\rangle)+ {1 \over \sqrt{2}}|1\rangle{1 \over \sqrt{2}}(|E_{0}\rangle-i|E_{1}\rangle).
\end{equation}
This time, when the ancilla qubit is measured as $0$, the expectation value of $Q$
is given by
\begin{equation}
\langle Q \rangle_{i0}={1 \over 2}(\langle E_{0}|Q|E_{0}\rangle+\langle E_{1}|Q|E_{1}\rangle-i\langle E_{1}|Q|E_{0}\rangle+i\langle E_{0}|Q|E_{1}\rangle).
\end{equation}
When the ancilla qubit is measured as $1$, the expectation value of $Q$
is given by
\begin{equation}
\langle Q \rangle_{i1}={1 \over 2}(\langle E_{0}|Q|E_{0}\rangle+\langle E_{1}|Q|E_{1}\rangle+i\langle E_{1}|Q|E_{0}\rangle-i\langle E_{0}|Q|E_{1}\rangle).
\end{equation}
By taking the difference between the Eqs. (11) and (12), we obtain
\begin{equation}
\langle Q \rangle_{i0}-\langle Q \rangle_{i1}=-i\langle E_{1}|Q|E_{0}\rangle+i\langle E_{0}|Q|E_{1}\rangle.
\end{equation}
From Eq. (13), the imaginary part of  $\langle E_{1}|Q|E_{0}\rangle$ is determined.
Thus, combining Eqs. (8) and (13) we can compute the matrix elements $\langle E_{1}|Q|E_{0}\rangle$ and $\langle E_{0}|Q|E_{1}\rangle$.

\section{A Single-qubit model with a single parameter}
In this section we consider the following single-qubit Hamiltonian parametrized by a single variable
\begin{equation}
{\hat H}=X+JZ,
\end{equation}
where $J$ is a real parameter and $X$ and $Z$ are Pauli matrices
\begin{equation}
X=\left(\begin{array}{cc} 0 & 1 \\ 1 & 0 \end{array}\right), \quad Z=\left(\begin{array}{cc} 1 & 0 \\ 0 & -1 \end{array}\right).
\end{equation}
Algebraic calculations show that the ground state energy is $E_{0}=-\sqrt{1+J^{2}}$ and the corresponding eigenstate is
\begin{equation}
|E_{0}\rangle={1 \over \sqrt{2}\sqrt{1+J^{2}+2J\sqrt{1+J^{2}}}}(|0\rangle-(J+\sqrt{1+J^{2}})|1\rangle).
\end{equation}
The excited state energy is $E_{1}=\sqrt{1+J^{2}}$ and the corresponding eigenstate is
\begin{equation}
|E_{1}\rangle={1 \over \sqrt{2}\sqrt{1+J^{2}-2J\sqrt{1+J^{2}}}}(|0\rangle+(-J+\sqrt{1+J^{2}})|1\rangle).
\end{equation}
The unitary operator $U_{10}$ that maps the ground state $|E_{0}\rangle$ to the excited state is given by
\begin{equation}
U_{10}=|E_{1}\rangle\langle E_{0}|+|E_{0}\rangle\langle E_{1}|.
\end{equation}
Using the algebraic expressions in Eqs. (16) and (17) for $J=1$,  the unitary operator $U_{10}(J=1)$ is given by
\begin{equation}
U_{10}(J=1)=|E_{1}\rangle\langle E_{0}|+|E_{0}\rangle\langle E_{1}|={1 \over \sqrt{2}}\left(\begin{array}{cc} 1 & -1 \\ -1 & -1 \end{array}\right)=ZH_{d}Z.
\end{equation}

Recently, we demonstrated that the ground state $|E_{0}\rangle$ and the excited state $|E_{1}\rangle$ can be generated using classically emulated
digital quantum simulation based on the Hamiltonian\cite{Oshima1}.   First, we  perform semi-algebraic simulations using Eq. (19). 
By operating the quantum circuits in Figs. 2 and 3 on the ground state generated via the classically emulated
digital quantum simulation using the quantum gate in Eq. (19) and performing appropriate measurements, we obtain the matrix elements of the form $\langle E_{0}|Q|E_{1}\rangle+\langle E_{1}|Q|E_{0}\rangle$ and $i(\langle E_{0}|Q|E_{1}\rangle-\langle E_{1}|Q|E_{0}\rangle)$ for a physical observable $Q$.   In Tables I and II we present our simulation results obtained using $U_{10}(J=1)=ZH_{d}Z$ (simulation partially assisted  by the algebraic expression). 
\\ \\ \\
Table I.   Simulation results of $\langle E_{0}|Q|E_{1}\rangle+\langle E_{1}|Q|E_{0}\rangle$  for $J=1$ .  The number of trials is $10^{7}$, and $U_{10}=ZH_{d}Z$ is used.
$Y=\left(\begin{array}{cc} 0 & -i \\ i & 0 \end{array}\right)$ is one of the Pauli matrices.\\
\\
\begin{tabular}{|c|c|c|c|}\hline
$Q$ & $X$ & $Z$ &  $Y$ \\ \hline
algebraic & $-\sqrt{2}$ & $\sqrt{2}$ & 0 \\ \hline
$U_{10}=ZH_{d}Z$ & -1.41360 &1.41327&  0.00011\\ \hline 
\end{tabular}  \\
\\ \\ \\
Table II.   Simulation results of $i(\langle E_{0}|Q|E_{1}\rangle-\langle E_{1}|Q|E_{0}\rangle)$ for $J=1$.   The number of trials is $10^{7}$, and $U_{10}=ZH_{d}Z$ is used.\\
\\
\begin{tabular}{|c|c|c|c|}\hline
$Q$ & $X$ & $Z$ &  $Y$ \\ \hline
algebraic & $0$ & $0$ & 2 \\ \hline
$U_{10}=ZH_{d}Z$ & -0.00057 &-0.00034&  2.00000\\ \hline 
\end{tabular}
\\
{} \qquad \qquad \\
{} \\ \\

The ground state $|E_{0}\rangle$ and the excited state $|E_{1}\rangle$ can be generated via a classically emulated
digital quantum simulation based on the Hamiltonian\cite{Oshima1}.  Thus, without requiring prior knowledge of the algebraic forms of $|E_{0}\rangle$ and $|E_{1}\rangle$,
we can construct the unitary operator $U_{10}$ using $|E_{0}\rangle$ and $|E_{1}\rangle$ obtained from simulations.
In the following, we determine the matrix representation of $U_{10}(J=1)$ based on the simulation results.

First, let us generate the ground state $|E_{0}\rangle$.  Starting from the initial physical state $|1\rangle$ and applying the four twirling operations\cite{Oshima1},
we obtain a good approximation to the ground state $|E_{0}\rangle$.   
We consider the two cases of $J=1$ and $J=2$.
In a simulation for $J=1$,  an approximate value of $\langle E_{0}|{ H}|E_{0}\rangle$ is -1.41413, $\langle E_{0}|Z|E_{0}\rangle=-0.70692$ and  $\langle E_{0}|X|E_{0}\rangle=-0.70721$ for $10^{7}$ trials.    From the latter two values we can estimate $|E_{0}\rangle$.   We set $|E_{0}\rangle$ as $|E_{0}\rangle={}^{t}(a, b{e^{i\phi}})$, where $a,b \ge 0$, we find that $a=0.3828$, $b=0.9238$, and $\cos\phi=-0.9999$. 
Since the Hamiltonian ${H}=X+JZ$ is Hermitian and its components are real, we can choose the eigenvectors $|E_{0}\rangle$ and $|E_{1}\rangle$ as real vectors. 
Therefore, we assume that $\cos\phi=-1$, and we conclude that the ground state for $J=1$ is $|E_{0}(J=1)\rangle={}^{t}(0.3828, -0.9238)$, which is in good agreement with Eq. (16) for $J=1$.

Next, let us generate the excited state $|E_{1}\rangle$.   By appropriately choosing the initial physical state, it is possible to arrive at the other energy eigenstate by the adaptive twirling operation.   Starting from the initial physical state $|0\rangle$, and applying the four twirling operations\cite{Oshima1}
we obtain a good approximation of the excited state $|E_{1}\rangle$.   
In a simulation for $J=1$, we obtain the following approximate values $\langle E_{1}|{ H}|E_{1}\rangle=1.41387$, $\langle E_{1}|Z|E_{1}\rangle=0.70700$ and  $\langle E_{1}|X|E_{1}\rangle=0.70687$  for
$10^{7}$ trials.   In the same way as in the case of $|E_{0}(J=1)\rangle$, we conclude that the excited state for $J=1$ is  $|E_{1}(J=1)\rangle={}^{t}(0.9238, 0.3828)$.

For $J=1$, using the above simulation results $|E_{0}(J=1)\rangle={}^{t}(0.3828, -0.9238)$ and  $|E_{1}(J=1)\rangle={}^{t}(0.9238, 0.3828)$, we obtain
\begin{equation}
U_{10}(J=1)=\left(\begin{array}{cc} 0.70729 & -0.70690 \\ -0.70690 & -0.70735 \end{array}\right),
\end{equation}
which is a simulation result that corresponds to Eq. (19).
In the next phase of simulation, the matrix above is not recognized as unitary in Qiskit, resulting in an error.   Instead, we use the following matrix, which has been obtained by unitarizing Eq. (20) using SciPy, a Python library:
\begin{equation}
U_{10}(J=1)=\left(\begin{array}{cc} 0.70731675 & -0.70689675 \\ -0.70689675 & -0.70731675 \end{array}\right) .
\end{equation}
To evaluate the accuracy of this approximation, we 
apply it to the above ground state $|E_{0}\rangle$ in a simulation to obtain an approximation of the excited state $|E_{1}\rangle$.   A simulation value obtained from the state $|E_{0}\rangle$ that satisfies $\langle E_{0}|{H}|E_{0}\rangle=-1.413922$, yields $\langle E_{1}|{H}|E_{1}\rangle=1.413957$.  This indicates
that the excited state $|E_{1}\rangle$ has been generated with considerable accuracy.    Furthermore, by applying the operator $U_{10}(J=1)$ to the excited state $|E_{1}\rangle$ in a simulation,  we can also obtain an approximate ground state $|E_{0}\rangle$.

By operating the quantum circuits in Figs 2 and 3 using the quantum gate  $U=U_{10}(J=1)$ in Eq. (21) and performing appropriate measurements, we can obtain the matrix elements $\langle E_{0}|Q|E_{1}\rangle+\langle E_{1}|Q|E_{0}\rangle$ and $i(\langle E_{0}|Q|E_{1}\rangle-\langle E_{1}|Q|E_{0}\rangle)$ for a physical observable $Q$.   In Tables III and IV we present our simulation results. 
The matrix in Eq. (21) appears to be sufficiently accurate within the range used in the simulation, even though Eq. (19) gives the exact result.
\\ \\ \\
Table III   Simulation results of $\langle E_{0}|Q|E_{1}\rangle+\langle E_{1}|Q|E_{0}\rangle$ for $J=1$ based on Eq. (21).   Each simulation was performed with $10^{7}$ trials.\\
\\
\begin{tabular}{|c|c|c|c|}\hline
$Q$ & $X$ & $Z$ &  $Y$ \\ \hline
algebraic & $-\sqrt{2}$ & $\sqrt{2}$ & 0 \\ \hline
simulation& -1.41326 &1.41614&  -0.00011\\ \hline 
\end{tabular}  \\
\\ \\
Table  IV   Simulation results of $i(\langle E_{0}|Q|E_{1}\rangle-\langle E_{1}|Q|E_{0}\rangle)$ for $J=1$  based on Eq. (21).  Each simulation was performed with $10^{7}$ trials. \\
\\
\begin{tabular}{|c|c|c|c|}\hline
$Q$ & $X$ & $Z$ &  $Y$ \\ \hline
algebraic & $0$ & $0$ & 2 \\ \hline
simulation & 0.00087 &-0.00027&  2.00000\\ \hline 
\end{tabular}
{} \qquad 
\\ \\

As a next step, we examine the case where the parameter takes the value $J=2$.
Using the algebraic expression in Eqs. (16) and (17) for $J=2$,  the unitary operator $U_{10}(J=2)$ is given by
\begin{equation}
U_{10}(J=2)={1 \over \sqrt{5}}\left(\begin{array}{cc} 1 & -2 \\ -2 & -1 \end{array}\right).
\end{equation}

Analogously to the case $J=1$,  for $J=2$ we obtain approximate values $\langle E_{0}|{\hat H}|E_{0}\rangle=-2.23748$, 
$\langle E_{0}|Z|E_{0}\rangle=-0.895776$ and $\langle E_{0}|X|E_{0}\rangle=-0.4459324$ from $10^{6}$ trials.  We therefore conclude that  $|E_{0}(J=2)\rangle={}^{t}(0.2298, -0.9733)$.
In  the same way,  for $J=2$ we have approximate values $\langle E_{1}|{\hat H}|E_{1}\rangle=2.23485, \langle E_{1}|Z|E_{1}\rangle=0.89419$ and $\langle E_{1}|X|E_{1}\rangle=0.446576$ from
$10^{6}$ trials.    In this case, we obtain $\cos\phi=0.9975$, and we assume that $\cos\phi=1$. Accordingly, we conclude that  $|E_{1}(J=2)\rangle={}^{t}(0.9732, 0.2300)$.   
The corresponding expression to Eq. (22) in this case is 
\begin{equation}
U_{10}(J=2)=\left(\begin{array}{cc} 0.44748977 & -0.89428905 \\ -0.89428905 & -0.44748977 \end{array}\right) .
\end{equation}
\\ 
\\
The simulation results based on Eq. (23) are summarized in Tables V and VI.\\
\\
Table V    Simulation results for $\langle E_{0}|Q|E_{1}\rangle+\langle E_{1}|Q|E_{0}\rangle$ with $J=2$.  The number of trials is $10^{7}$, and Eq. (23) is used.    \
\\
\begin{tabular}{|c|c|c|c|}\hline
$Q$ & $X$ & $Z$ &  $Y$ \\ \hline
algebraic & $-{4 \over \sqrt{5}}$ & ${2 \over \sqrt{5}}$ & 0 \\ \hline
simulation &-1.78857 &0.89519&-0.00035\\ \hline 
\end{tabular}  \\
\\
\\
Table VI   Simulation results for $i(\langle E_{0}|Q|E_{1}\rangle-\langle E_{1}|Q|E_{0}\rangle)$ with $J=2$.    The number of trials is $10^{7}$, and Eq. (23) is used.  \\
\\
\begin{tabular}{|c|c|c|c|}\hline
$Q$ & $X$ & $Z$ &  $Y$ \\ \hline
algebraic & $0$ & $0$ & 2 \\ \hline
simulation & 0.00048 &0.00100&  2.00000\\ \hline 
\end{tabular}
{} \\ \\ \\
For $J=1$ and $J=2$, the simulated values of $\langle E_{1}|Q|E_{0}\rangle$ agree well with the corresponding algebraic calculations.
This indicates that the superposition states ${1 \over \sqrt{2}}(|E_{0} \rangle \pm |E_{1}\rangle)$ and  ${1 \over \sqrt{2}}(|E_{0} \rangle \pm i|E_{1}\rangle)$ have been constructed with sufficient accuracy through the simulation. 

\section{Hydrogen molecule}
In this section, we will consider the Hamiltonian of the hydrogen molecule, which plays a central role in quantum chemistry.   The Hamiltonian of a hydrogen molecule is given by \cite{Omalley, Colless, Ganzhorn}
\begin{equation}
H=a_{0}I_{0}I_{1}+a_{1}Z_{0}I_{1}+a_{2}I_{0}Z_{1}+a_{3}Z_{0}Z_{1}+a_{4}X_{0}X_{1},
\end{equation}
where $\{a_{i}\}$ are constants that depend on the internuclear distance, $I$, $X$ and $Z$ represent the identity operator and the Pauli matrices, and the
subscripts 0 and 1 on these matrices indicate that they act respectively on the 0th and 1st qubits of the quantum state.
The constants $\{a_{i}\}$ are given by
$a_{0}=-1.04319, a_{1}=-a_{2}=0.42045, a_{3}=0.01150, a_{4}=0.179005 [{\rm E_{h}}]$, where $1[{\rm E_{h}}]=27.211386[{\rm eV}]$, for an internuclear distance of 0.70 ${\rm \AA}$\cite{Singh}.   By means of algebraic computation, the energy eigenvalues of the Hamiltonian are derived in increasing order as follows:
\begin{equation}
E_{0}=-1.89215, \quad E_{1}=-1.23442, \quad E_{2}=-0.876405, \quad E_{3}=-0.172668[{\rm E_{h}}].
\end{equation}
The corresponding energy eigenstates are obtained algebraically in the computational basis $\{|0\rangle|0\rangle, |0\rangle|1\rangle, |1\rangle|0\rangle, |1\rangle|1\rangle\}$ as 
\begin{eqnarray}
|E_{0}\rangle=\left(\begin{array}{c} 0 \\ 0.104679 \\ -0.994506 \\0 \end{array}\right),
|E_{1}\rangle=\left(\begin{array}{c} -0.707107 \\ 0 \\ 0 \\0.707107 \end{array}\right),
|E_{2}\rangle=\left(\begin{array}{c} 0.707107 \\ 0 \\ 0 \\0.707107 \end{array}\right),
|E_{3}\rangle=\left(\begin{array}{c} 0 \\  0.994506\\0.104679 \\ 0 \end{array}\right).
\end{eqnarray} 
The following is a unitary operator that transforms the ground state $|E_{0}\rangle$ into the first excited state $|E_{1}\rangle$ obtained from the algebraic results in Eq. (26).
\begin{eqnarray}
U_{10{\rm alg}}&=&|E_{1}\rangle\langle E_{0}|+|E_{0}\rangle\langle E_{1}|+|E_{2}\rangle\langle E_{2}|+|E_{3}\rangle\langle E_{3}| \nonumber \\
&=&\left( \begin{array}{cccc} 0.5 & -0.074019 & 0.7032219 & 0.5 \\ -0.074019 & 0.989042 & 0.104104 & 0.074019 \\ 0.7032219 & 0.104104 & 0.01096 & -0.7032219\\
0.5 & 0.074019 & -0.7032219 & 0.5 \end{array} \right).
\end{eqnarray}
Since the matrix $U_{10{\rm alg}}$ is not recognized as unitary by Qiskit, we will use the following matrix $U_{10{\rm alg}^{\prime}}$ that has been obtained through the
polar decomposition of the matrix $U_{10{\rm alg}}$ using SciPy, a Python library.
\begin{eqnarray}
U_{10{\rm alg}^{\prime}}&=&\left( \begin{array}{cccc} 0.49999944& -0.07401925 & 0.70322198 & 0.500000056 \\ -0.07401925 & 0.98904231 & 0.10410381 & 0.07401925 \\ 0.70322198 & 0.104104381 & 0.01095882 & -0.70322198\\
0.500000056 & 0.07401925 & -0.70322198 & 0.49999944 \end{array} \right).
\end{eqnarray}
This matrix $U_{10{\rm alg}^{\prime}}$ is recognized as a unitary matrix by Qiskit.   Starting from the initial state $|1\rangle_{0}|0\rangle_{1}$, and applying the four twirling operations\cite{Oshima2}
we obtain a good approximation to the ground state $|E_{0}\rangle$.   In the simulation, an approximate value of $\langle E_{0}|H|E_{0}\rangle$ is -1.8925087 for
$10^{6}$ trials.   To this state, we apply the unitary matrix $U_{{\rm alg}^{\prime}}$ in Eq. (28) to obtain the first excited state $|E_{1}\rangle$.    Our simulation result of $\langle E_{1}|{H}|E_{1}\rangle$ is -1.234407.

First, we perform semi-algebraic simulations using Eq. (28). 
By operating the quantum circuits in Figs. 2 and 3 using the quantum gate $U_{10}=U_{10{\rm alg}^{\prime}}$ in Eq. (28) and performing appropriate measurements, we can obtain the matrix elements $\langle E_{0}|Q|E_{1}\rangle+\langle E_{1}|Q|E_{0}\rangle$ and $i(\langle E_{0}|Q|E_{1}\rangle-\langle E_{1}|Q|E_{0}\rangle)$ of the physical observable $Q$.   The simulation results are presented in Table VII
and Table VIII. 
\\ \\
Table VII  Simulation results of $\langle E_{1}|Q|E_{0}\rangle+\langle E_{0}|Q|E_{1}\rangle$ obtained by $10^{7}$ trials.      We have prepared an approximate ground state starting from $|1\rangle_{0}|0\rangle_{1}$ by four twirling operations.  The quantum circuit in Fig. 2 has been executed with $U_{10}=U_{10{\rm alg}^{\prime}}$ in Eq. (28).\\
\\
\begin{tabular}{|c|c|c|c|c|c|c|c|}\hline
$Q$ & $Z_{0} \otimes Z_{1}$ & $Z_{0} \otimes I_{1}$ &  $I_{0} \otimes Z_{1}$ &  $X_{0} \otimes X_{1}$  & $X_{0} \otimes I_{1}$ &  $Y_{0} \otimes Y_{1}$ &  $Y_{0} \otimes I_{1}$\\ \hline
algebraic & 0 & 0 & 0 & 0 & 1.5545 & 0 & 0\\ \hline
$U_{10alg^{\prime}}$ & -3.7$\times$ $10^{-6}$ &-0.00033 & -0.00042 & -0.0054 & 1.5542 & 0.0015 & 0.00078\\ \hline 
\end{tabular}
\\ \\ \\
Table VIII  Simulation results of $-i\langle E_{1}|Q|E_{0}\rangle+i\langle E_{0}|Q|E_{1}\rangle$  obtained by $10^{7}$ trials.   We have prepared an approximate ground state starting from $|1\rangle_{0}|0\rangle_{1}$ by four twirling operations.  The quantum circuit in Fig. 3 has been executed with $U_{10}=U_{10{\rm alg}^{\prime}}$ in Eq. (28).\\
\\
\begin{tabular}{|c|c|c|c|c|c|c|c|}\hline
$Q$ & $Z_{0} \otimes Z_{1}$ & $Z_{0} \otimes I_{1}$ &  $I_{0} \otimes Z_{1}$ &  $X_{0} \otimes X_{1}$  & $X_{0} \otimes I_{1}$ &  $Y_{0} \otimes Y_{1}$ &  $Y_{0} \otimes I_{1}$\\ \hline
algebraic & 0 &0 & 0 &0  & 0 & 0 & -1.2584\\ \hline
$U_{10alg^{\prime}}$ & 1.4$\times$ $10^{-7}$ &-0.00315 & 0.00206 & 0.00050 & -0.00091  & 0.00175 & -1.2580\\ \hline 
\end{tabular}
\\ \\ \\

Next, we perform simulations based solely on the Hamiltonian and the results of prior simulations, assuming no prior knowledge of algebraic outcomes.
In what follows,  we assume no prior algebraic knowledge of the eigenstates $|E_{i}\rangle$ for $i=0,1,2,3$.
As in the single-qubit model, the four energy eigenstates $|E_{i}\rangle(i=0,1,2,3)$ have been generated by four or six successive twirling operations
and six expectation values $\langle E_{i}|XX|E_{i}\rangle, \langle E_{i}|ZX|E_{i}\rangle, \langle E_{i}|XZ|E_{i}\rangle, $
$\langle E_{i}|ZI|E_{i}\rangle, \langle E_{i}|IZ|E_{i}\rangle$
and $\langle E_{i}|ZZ|E_{i}\rangle$ have been computed for each energy eigenstate $|E_{i}\rangle$.
\\ \\ \\
Table IX  Simulation results of $\langle E_{i}|Q|E_{i}\rangle$, where $i=0,1,2,3.$  A total of $10^{7}$ repetitions have been performed for each state.    The adaptive twirling operation was applied four times to generate
the states $|E_{0}\rangle$ and $|E_{3}\rangle$, and six times for the states $|E_{1}\rangle$ and $|E_{2}\rangle$. \\
\\
\begin{tabular}{|c|c|c|c|c|c|c|}\hline
$Q$ & $Z_{0} \otimes Z_{1}$ & $Z_{0} \otimes I_{1}$ &  $I_{0} \otimes Z_{1}$ &  $X_{0} \otimes X_{1}$  & $Z_{0} \otimes X_{1}$ &  $X_{0} \otimes Z_{1}$ \\ \hline
$|E_{0}\rangle$ & -0.9992030 & -0.9780797 & 0.9780797 & -0.2078699 & 0.0007952 & 0.0001650 \\ \hline
$|E_{1}\rangle$ & 1.0009224 &-0.0004632 & -0.0004366 & -1.0007123 & 0.0050201 & 0.0030851 \\ \hline 
$|E_{2}\rangle$  & 0.9963039 &-0.0024022 & -0.0013617 & 0.9993067 & -0.0210744 & -0.0210744 \\ \hline 
$|E_{3}\rangle$  & -0.9998782  &  0.9780336 & -0.9780336 &    0.2082828  &0.000260      & 0.0002596  \\ \hline
\end{tabular}
{} \qquad \\ \\
\\ \\
The states $|E_{i}^{sim}\rangle$ for $i=0,1,2,3$ are estimated by Python's SciPy library as the most plausible states consistent with the data in Table IX,    They are obtained as follows:
\begin{eqnarray}
|E_{0}^{sim}\rangle=\left(\begin{array}{c} -0.000126 \\ 0.104525 \\ -0.994522 \\0.000408 \end{array}\right),
|E_{1}^{sim}\rangle=\left(\begin{array}{c} -0.706940 \\ -0.001111 \\ -0.001752 \\0.707271 \end{array}\right), 
|E_{2}^{sim}\rangle=\left(\begin{array}{c} 0.706598 \\ -0.037866 \\ -0.023105 \\0.706223 \end{array}\right), 
|E_{3}^{sim}\rangle=\left(\begin{array}{c} 0.000133 \\  0.994501\\0.104728 \\ -0.000022 \end{array}\right).
\end{eqnarray}
\\
An approximate matrix representation of the unitary operator $U_{10}$ is derived from the simulation results in Eq. (29)
\\
\begin{eqnarray}
U_{10}^{sim}&=&|E_{1}^{sim}\rangle\langle E_{0}^{sim}|+|E_{0}^{sim}\rangle\langle E_{1}^{sim}|+|E_{2}^{sim}\rangle\langle E_{2}^{sim}|+|E_{3}^{sim}\rangle\langle E_{3}^{sim}| \nonumber \\
&=&\left( \begin{array}{cccc} 0.499457 & -0.1005 & 0.686603 & 0.49864 \\ -0.1005 & 0.989734 & 0.108495 & 0.0472221 \\ 0.686603 & 0.108495 & 0.0117281 & -0.719858\\
0.49864 & 0.047221 & -0.719858 & 0.499328 \end{array} \right).
\end{eqnarray}
\\
Using the Python library SciPy, Eq. (30) is replaced with the following unitary operator
\\
\begin{eqnarray}
U_{10}^{sim^{\prime}}=\left( \begin{array}{cccc} 0.50997726 & -0.07457948 & 0.69620855 & 0.49965463 \\ -0.07457945 & 0.98864935 & 0.10595974 & 0.07604568 \\
 0.69620856 & 0.10596012 & 0.01084928 & -0.70989323\\
0.49965467 & 0.07604487 & -0.70989329 & 0.49052412 \end{array} \right).
\end{eqnarray}
Using this unitary matrix, we construct the superposition state ${1 \over \sqrt{2}}(|E_{0}\rangle+|E_{1}\rangle)$ with the ancilla qubit in the state $|0\rangle$ 
and the superposition state ${1 \over \sqrt{2}}(|E_{0}\rangle-|E_{1}\rangle)$ with the ancilla qubit in the state $|1\rangle$ in the quantum circuit shown in Fig. 2.
In addition, we construct the superposition state ${1 \over \sqrt{2}}(|E_{0}\rangle+i|E_{1}\rangle)$ with the ancilla qubit in the state $|0\rangle$ 
and the superposition state ${1 \over \sqrt{2}}(|E_{0}\rangle-i|E_{1}\rangle)$ with the ancilla qubit in the state $|1\rangle$ in the quantum circuit shown in Fig. 3.
The simulation results are presented in Tables X and XI.
\\
\\
Table X  Simulation results for $\langle E_{1}|Q|E_{0}\rangle+\langle E_{0}|Q|E_{1}\rangle$.   The entries labeled by "$U_{10{\rm alg}^{\prime}}$" were tested over $10^{7}$ trials using Eq. (28).
The entries labeled by "simulation" represent the average of ten runs, each consisting of $10^{7}$ trials.    The 95\% confidence interval is shown in the row below.
\\
\\
\begin{tabular}{|c|c|c|c|c|c|c|c|}\hline
$Q$ & $Z_{0} \otimes Z_{1}$ & $Z_{0} \otimes I_{1}$ &  $I_{0} \otimes Z_{1}$ &  $X_{0} \otimes X_{1}$  & $X_{0} \otimes I_{1}$ &  $Y_{0} \otimes Y_{1}$ &  $Y_{0} \otimes I_{1}$\\ \hline
algebraic & 0 & 0 & 0 & 0 & 1.5545 & 0 & 0\\ \hline
 $U_{10alg^{\prime}}$ & $-3.7\times 10^{-6}$ &-0.00033 & -0.00042 & -0.0054 & 1.5542 & 0.0015 & 0.00078\\ \hline 
simulation & $-3.1\times 10^{-7}$ &-0.00022 & -0.00026 & 0.0031 & 1.5422 & 0.0040 & 0.00042\\ \hline 
95 $\%$ interval &  $-3.1\times 10^{-7}$   &  0.00032  & 0.00036 &    0.00069  &0.00026      & 0.00040 &  0.00038  \\ \hline
\end{tabular}
\\ \\ \\
Table XI  Simulation results for $-i\langle E_{1}|Q|E_{0}\rangle+i\langle E_{0}|Q|E_{1}\rangle$.   The entries labeled by "$U_{10{\rm alg}^{\prime}}$" have been tested with $10^{7}$ trials  using Eq. (28).
The entries labeled by "simulation" represent the average of ten runs, each consisting of $10^{7}$ trials.    The 95\% confidence interval is shown in the row below.\\
\\
\begin{tabular}{|c|c|c|c|c|c|c|c|}\hline
$Q$ & $Z_{0} \otimes Z_{1}$ & $Z_{0} \otimes I_{1}$ &  $I_{0} \otimes Z_{1}$ &  $X_{0} \otimes X_{1}$  & $X_{0} \otimes I_{1}$ &  $Y_{0} \otimes Y_{1}$ &  $Y_{0} \otimes I_{1}$\\ \hline
algebraic & 0 &0 & 0 &0  & 0 & 0 & -1.2584\\ \hline
$U_{10{\rm alg}^{\prime}}$ & $1.4\times 10^{-7}$ &-0.00315 & 0.00206 & 0.00050 & -0.00091  & 0.00175 & -1.2580\\ \hline
simulation & $-5.8\times 10^{-8}$ &0.00025 & $ 5.8\times 10^{-5}$ & 0.00014 & -0.0013  & 0.00011 & -1.2433\\ \hline 
95$\%$ confidence interval &  $1.4\times 10^{-7}$   &  0.00044  & 0.00041 &    0.00047  &0.00072      & 0.00052 &  0.00029  \\ \hline
\end{tabular}
\\ \\ \\ \\
The values of the matrix elements $\langle E_{1}|Q|E_{0}\rangle+\langle E_{0}|Q|E_{1}\rangle$ and $-i\langle E_{1}|Q|E_{0}\rangle+i\langle E_{0}|Q|E_{1}\rangle$, labeled as "simulation" in Tables X and XI calculated from the quantum states generated by the circuits in Figs. 2 and 3, closely match those obtained through algebraic calculations.    Therefore, the quantum states stored in the two quantum circuits can be regarded as the desired superposition states.\\

\section{Summary and discussions}

This work extends our previously proposed method for generating energy eigenstates.
By employing a unitary operator that maps the ground state of the hydrogen molecule to its first excited state, we have generated superpositions of the ground state and the first excited state through
classically emulated quantum computation.    Based on the expectation values of several physical observables, these superposition states appear to be constructed with reasonably satisfactory accuracy.
In constructing the superposition states, we considered both the case in which no algebraic results were used and the case in which such results were incorporated 
into the raising unitary operator.    Our method is applicable not only to the construction of superpositions involving the ground and first excited states, but also to those comprising arbitrary
energy eigenstates.    As a preliminary step toward constructing superpositions of energy eigenstates of the hydrogen molecule, we first considered a simple Hamiltonian $H=X+JZ$, characterized by a single parameter $J$.    In principle, our method is applicable to arbitrary Hamiltonians expressible in terms of Pauli operators. 

\section{Acknowledgements}
The author thanks National Institute of Technology, Gunma College, where this study was conducted.

\newpage
Figure captions\\
\\
Fig. 1. Twirling operation for generating an energy eigenstate.    The choice of duration $\tau$ determines which energy eigenstate's amplitude is enhanced.\\
\\
\\
Fig. 2. Quantum circuit for constructing the superposition states ${1 \over \sqrt{2}}(|E_{0}\rangle{\pm}|E_{1}\rangle)$ starting from the ground state $|E_{0}\rangle$.\\
\\
\\
Fig. 3. Quantum circuit for constructing the superposition states ${1 \over \sqrt{2}}(|E_{0}\rangle{\pm}i|E_{1}\rangle)$ starting from the ground state $|E_{0}\rangle$.\\

\newpage
{\quad } \includegraphics[width=8cm]{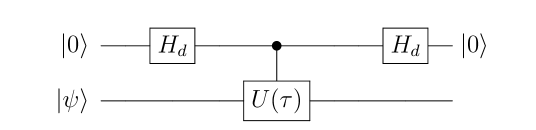}
\\
Fig. 1\\
\\
\vspace{10pt}
{\quad }  \includegraphics[width=8cm]{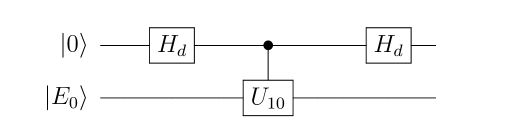}
\\
Fig. 2\\
\vspace{10pt}
\\ \\ \\
{\qquad}  \includegraphics[width=10cm]{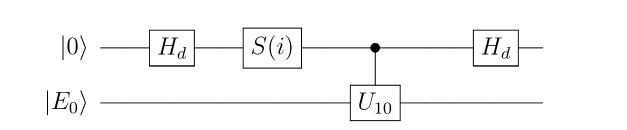}
\\
Fig. 3\\
\end{document}